\documentclass[aps,nofootinbib,preprintnumbers,showpacs,prd,twocolumn,superscriptaddress]{revtex4-1}

\usepackage{mathrsfs}
\usepackage{graphicx}
\usepackage{float}
\usepackage{bm}
\usepackage{hyperref}
\usepackage{amsmath}
\usepackage{amssymb}
\usepackage{lipsum}
\begin{document}

\title{Quark Wigner distribution of the pion meson in light-cone quark model}
\author{Zhi-Lei Ma}
\affiliation{Department of Physics, Yunnan University, Kunming 650091, China}
\author{Zhun Lu}
\email{zhunlu@seu.edu.cn}
\affiliation{School of Physics, Southeast University, Nanjing 211189, China}

\begin{abstract}
We study the quark Wigner distributions of the pion meson which are the phase-space distributions containing the most general one-parton information.
Using the pion wavefunctions deduced from a light-cone quark model, we calculate the Wigner distributions of the unpolarized, longitudinally polarized and transversely polarized quarks inside the pion meson via the Fock-state overlap representation, respectively.
We present the numerical results of the transverse Wigner distributions in which the longitudinal momentum fraction is integrated out.
The mixed Wigner distributions as functions of $b_y$ and $k_x$ are also presented.
As a intermediate step, we provide the analytic results for the generalized transverse momentum dependent distributions of the pion meson in the same model.
\end{abstract}

\maketitle

\section{Introduction}
\label{Sec.intro}

Understanding the internal structures of hadrons in terms of quarks and gluons and their roles in high energy scattering process is one of the main goals of QCD and hadronic physics.
Parton distribution function (PDF) $f_{i/h}(x)$, which expresses the probability density of parton $i$ in a hadron $h$ with the longitudinal momentum fraction $x$, was first introduced by Feynman~\cite{Feynman:1969ej} to describe inclusive process with one hadron in the initial state. Although very successful, PDFs can only describe partonic structure of hadrons in one dimension.
In the last two decades, a much more comprehensive picture on the nucleon structure has been developed~\cite{Mulders:1995dh,Goeke:2005hb,Bacchetta:2006tn}, and the transverse momentum dependent (TMD) distributions play the central role.
TMD distributions $f(x,\bm k_\perp)$ not only depend on the longitudinal momentum fraction $x$, but also depend on the parton transverse momentum $\bm k_\perp$ with respect to the hadron, therefore they allows a three-dimensional description of parton structure in momentum space.
TMD distributions naturally enter the description of semi-inclusive deep inelastic scattering and Drell-Yan process in which two hadrons are involved.
Furthermore, Combining the spin of the nucleon or the parton, TMD distributions unravel some novel features such as the Sivers effect~\cite{Sivers:1989cc,Sivers:1990fh} and the Boer-Mulders effect~\cite{Boer:1997nt} through spin-orbit correlation~\cite{Brodsky:2002cx,Collins:2002kn,Ji:2002aa}, which cannot be explained within the leading-twist collinear picture.

On the other hand, in the off-forward region a new type of nucleon structure--the so-call generalized parton distribution (GPD)~\cite{Mueller:1998fv,Ji:1996nm,Radyushkin:1997ki}--emerges.
It is the extension of the ordinary PDF from the forward scattering region to the off-forward scattering region.
Therefore, GPDs are the natural observables appearing in various exclusive processes in which the target receives a recoil momentum $\Delta$, such as the deeply virtual Compton scattering (DVCS) $\gamma^* h(p) \rightarrow\gamma h (p^\prime)$ and the hard exclusive production of meson $\gamma^* h_1(p) \rightarrow M h_2(p^\prime)$.
Except $x$, GPDs also depend on the momentum transfer squared $t=\Delta^2$ and the longitudinal fraction $\xi=\Delta^+ /P^+$ of the transferred momentum. GPDs may reduce to various form factors after $x$ is integrated out.
Particularly, Fourier transforming GPDs with respect to the transverse component of $\Delta$ yields the impact-parameter dependent distributions (IPDs)~\cite{Burkardt:2000za,Burkardt:2002hr,Diehl:2002he}, $f(x,\bm b_\perp^2)$, with $b_\perp$ the impact parameter conjugate to $\Delta_\perp$. IPDs thus
provide very useful information on the parton tomography in hadrons: distributions of parton in the transverse coordinate space at a fixed $x$.

A more fundamental understanding of the partonic structure of the nucleon can be gained by unifying the distributions in momentum space and in position space.
The Wigner distributions, which are functions of the full phase-space variables $\bm k$ and $\bm r$, can achieve this goal.
They provides most precise and complete description for the parton information in a hadron.
The six-dimensional (3 dimensions in position and 3 dimension in momentum) Wigner distributions in QCD were first introduced in Ref.~\cite{Ji:2003ak,Belitsky:2003nz}.
A very useful phase-space distribution for describing a fast moving hadron (or in the infinite-momentum frame) is the five-dimensional Wigner distribution~\cite{Lorce:2011kd}, denoted by $W(x,\bm k_\perp, \bm b_\perp)$.
In this case the light-cone formalism is a suitable approach, and in leading-twist the Wigner distribution can be expressed as the overlap integration of hadronic light-cone wavefunctions~\cite{Brodsky:1997de}.
After the transverse coordinates or the transverse momenta are integrated out,
the five-dimensional Wigner distributions reduce to the normal TMDs or IPDs, respectively.
However, in some cases there are no corresponding TMDs or IPDs for certain Wigner distributions after the spin of the hadron or the quark is considered.
Therefore the Wigner distributions also provide unique information that cannot be gained from the conventional three-dimensional distributions.
Furthermore, the Wigner distributions usually serve as the starting point for calculating other physical observables, especially the quark orbital angular momentum~\cite{Lorce:2011ni}.
Finally, the Wigner distributions can be also related to the generalized transverse momentum dependent distributions (GTMDs) through Fourier transformation.


In recent years, the Wigner distributions of the nucleon, for both the quark~\cite{Lorce:2011kd,Liu:2015eqa,Mukherjee:2014nya} and gluon~\cite{Mukherjee:2015aja,More:2017zqp} with different polarization configurations, have been calculated by various models.
There are also studies on the feasibility to probe the gluon Wigner distributions~\cite{Hatta:2016dxp,Hatta:2017cte,Hagiwara:2017fye}
as well as the GTMDs of the nucleon~\cite{Bhattacharya:2017bvs} through experimental measurements.
However, the Wigner distribution of the pion meson was less studied in literature.
So far only the GTMDs of the pion meson were calculated in a spectator model~\cite{Meissner:2008ay}.
This motivate us to present a first calculation on the quark Wigner distributions of the pion meson, which can provide complimentary information on the multidimensional description of the partonic structure of hadrons.
We will focus on the five-dimensional Wigner distributions $\rho(x,\bm b_\perp, \bm k_\perp)$, for which the light-cone formalism is a proper approach in the study.
Thus we will apply a light-cone quark model for the pion meson~\cite{Xiao:2003wf}.
In the calculation we consider the Wigner distributions for the unpolarized quark, longitudinally polarized quark and transversely polarized quark.
It is also interesting to compare the quark Wigner distributions of the pion meson with those of the nucleon
to find the common feature and the difference between them.

The rest of the paper is organized as follows. In Section.~II we review the systematics of the quark Wigner distributions of the pion and their connection with GTMDs. In Section.~III we derive the analytic expressions of the five-dimensional Wigner distributions of the pion meson in an overlap representation of the light-cone formalism. In Section.~IV we present the numerical results of the transverse Winger distributions as well as the mixed Wigner distribution. We summarize the paper in Section V.

\section{Systematics of quark Wigner distributions of pion meson}

Our starting point is the generalized quark-quark correlator for a spin-0 hadron introduced in Refs.~\cite{Meissner:2008ay,Meissner:2009ww}
\begin{align}
&W_{ij}(x, P, k_\perp)  \nonumber\\
&=
 \int \frac{dz^- d \bm z_\perp}{(2 \pi)^3} \, e^{i k \cdot z} \,
 \langle p^{\prime} \, | \, \bar{\psi}_j(-\tfrac{1}{2}z) \,
 {\cal W} \,
 \psi_i(\tfrac{1}{2}z) \, | \, p \rangle \Big|_{z^+ = 0}\, ,\label{eq:corr_gpcf}
\end{align}
with parton transverse momenta fully-unintegrated.
Here, $P = (p+p^\prime)/2$  is the averaged momentum of the
initial and final hadron, $\Delta = p^{\prime} - p$ is the the momentum transfer to the hadron, $k$ and  $x= k^+/P^+$  are the average momentum and the average longitudinal momentum fraction carried by the active quark.
To ensures the color gauge invariance of the correlator in Eq.~(\ref{eq:corr_gpcf}), the Wilson line ${\cal W}$ running along the path
\begin{equation} \label{e:path}
-\tfrac{1}{2}z \;\to\; -\tfrac{1}{2}z + \infty \cdot n
\;\to\; \tfrac{1}{2}z + \infty \cdot n \;\to\; \tfrac{1}{2}z \,,
\end{equation}
is included, with all four points connected by straight lines.
In Eq.~\ref{eq:corr_gpcf} we have suppressed two arguments of the correlator $W$: $N=M^2 n/P\cdot n$  and $\eta = \textrm{sign}(n_0)$, as they will not affect our discussion.

The correlator in Ref.~\ref{eq:corr_gpcf} can be parameterized into the GTMDs, on the basis of Dirac structure $\Gamma$:
\begin{align}
&W^{\Gamma}(x, P, k_\perp)  = {1\over 2} \textrm{Tr}
[W(P, x, k_\perp)\Gamma]\nonumber\\
& \qquad = \int \frac{dz^- \, d^2 \bm z_\perp}{2 (2\pi)^3} \, e^{i k \cdot z} \,
 \langle p^{\prime} \, | \, \bar{\psi}(-\tfrac{1}{2}z) \,
 \Gamma \, {\cal W} \,
 \psi(\tfrac{1}{2}z) \, | \, p \rangle\, \Big|_{z^+ = 0} \,.
 \label{eq:gcol}
\end{align}
Here we only consider the leading-twist decomposition.
For an unpolarized hadron, the parameterization has the form
\begin{align}
W^{[\gamma^+]} & =  F_{1,1} \,, \vphantom{\frac{1}{1}}
\label{eq:f11} \\
W^{[\gamma^+ \gamma_5]} & =
 -\frac{i\varepsilon_\perp^{ij} k_\perp^i \Delta_\perp^j}{M^2} \, G_{1,1} \,,
\label{eq:g11}\\
W^{[i\sigma^{j+}\gamma_5]} & =
 -\frac{i\varepsilon_\perp^{ij} k_\perp^i}{M} \, H_{1,1}
-\frac{i\varepsilon_\perp^{ij} \Delta_\perp^i}{M} \, H_{1,2} \,,
\label{eq:h1}
\end{align}
where $F_{1,1}$, $G_{1,1}$, $H_{1,1}$ and $H_{1,2}$ are the twist-2 GTMDs $X(x, \xi,\bm k_\perp^2, \bm k_\perp\cdot\bm\Delta_\perp, \bm \Delta_\perp^2)$ depending on the set of variables $x, \xi,\bm k_\perp^2, \bm k_\perp\cdot\bm \Delta_\perp, \bm \Delta_\perp^2$, with $\xi=\Delta^+/P^+$ the longitudinal momentum fraction of the transferred momentum and $\bm \Delta_\perp$ the transverse component of $\Delta$.
$M$ is the mass of the pion, $\varepsilon_\perp^{ij}$ is the anti-symmetric tensor in the transverse plane, and $\sigma^{j+} = {i\over 2}[\gamma^j, \gamma^+]$ is the Dirac tensor matrix.

Taking the $\Delta=0$ limit, the GTMDs can be projected to the TMD distributions appearing in SIDIS and Drell-Yan process. Integrating out the transverse momentum, the GTMDs reduces to the GPDs describing the exclusive processes.
Particlarly, there is no corresponding TMD or GPD for the GTMD $G_{1,1}$, therefore, it provides unique information that cannot be obtained from the conventional three-dimensional distributions
Furthermore, the GTMDs can be used to define the Wigner distributions of partons inside hadron through the Fourier transfer with respect to $\bm \Delta_\perp$
\begin{equation}
\rho^{[\Gamma]}(x,\bm b_\perp,\bm k_\perp)=\int\frac{\mathrm{d}^2\Delta_\perp}{(2\pi)^2}\,e^{-i\bm\Delta_\perp\cdot\bm b_\perp}\,W^{[\Gamma]}(x,\bm\Delta_\perp,\bm k_\perp),\label{eq:wigner1}
\end{equation}
where $W^{[\Gamma]}(x,\bm\Delta_\perp,\bm k_\perp)$ is the generalized correlator at $\xi=0$, and $\bm b_\perp$ is the impact parameter in the position space conjugate to $\bm \Delta_\perp$.
Particularly,
\begin{align}
\rho_{UU}(x,\bm b_\perp,\bm k_\perp)&=\rho^{[\gamma^+]}(x,\bm b_\perp,\bm k_\perp), \\
\rho_{UL}(x,\bm b_\perp,\bm k_\perp)&=\rho^{[\gamma^+\gamma_5]}(x,\bm b_\perp,\bm k_\perp),\\
\rho_{UT}(x,\bm b_\perp,\bm k_\perp)&=\rho^{[i\sigma^{j+}\gamma_5]}(x,\bm b_\perp,\bm k_\perp),
\end{align}
represent the Wigner distributions of the unpolarized quark, longitudinally polarized quark and transversely polarized quark in an unpolarized hadron, respectively.

According to Refs.~\cite{Ji:2003ak,Belitsky:2003nz}, the Wigner distributions can be also deduced from the Wigner operators for quarks at a fixed light-cone time $y^+=0$ as follows
\begin{align}\label{wigner-operator}
\widehat W^{[\Gamma]}(x,\bm b_\perp,\bm k_\perp)&=\frac{1}{2}\int\frac{\mathrm{d} z^-\,\mathrm{d}^2z_\perp}{(2\pi)^3}\,e^{i(k\cdot\bm z)}\nonumber\\
&\times\bar{\psi}(y-\tfrac{z}{2})\Gamma\mathcal W\,\psi(y+\tfrac{z}{2})\big|_{z^+=0},
\end{align}
with $y^\mu=[0,0,\bm b_\perp]$. Sandwiching the Wigner operators between the initial- and final-state of the nucleon yields the Wigner distribution~\cite{Lorce:2011kd}
\begin{align}
&\rho^{[\Gamma]}(x,\bm b_\perp,\bm k_\perp)\nonumber\\
\equiv&\int\frac{\mathrm{d}^2\Delta_\perp}{(2\pi)^2}\,\langle p^+,\tfrac{\bm\Delta_\perp}{2}|\widehat W^{[\Gamma]}(x,\bm b_\perp,\bm k_\perp)|p^+,-\tfrac{\bm\Delta_\perp}{2}\rangle.
\label{eq:wigner2}
\end{align}
which is equivalent to the definition in Eq.~\ref{eq:wigner1}.
In this picture, the variables $\bm b_\perp$ and $\bm k_\perp$ can be interpreted~\cite{Lorce:2011kd} as the relative average transverse position and the relative average transverse momentum of the quark, respectively.

Finally, the quark Wigner distributions can be related to the corresponding GTMDs~\cite{Lorce:2011kd}
\begin{align}
\rho_{UU}& = \mathcal{F}_{1,1}(x, 0,\bm k_\perp^2, \bm k_\perp\cdot\bm b_\perp, \bm b_\perp^2), \\
\rho_{UL}& = \frac{\epsilon_\perp^{ij}}{M^2}\bm k_{\perp\,i}{\partial\over \partial{\bm{b}_{\perp}^j}}\mathcal{G}_{1,1}(x, 0,\bm k_\perp^2, \bm k_\perp\cdot\bm b_\perp, \bm b_\perp^2), \\
\rho_{UT} &=\frac{\epsilon_\perp^{ij}}{2M}\bm k_{\perp\,i} \mathcal{H}_{1,1}(x, 0,\bm k_\perp^2, \bm k_\perp\cdot\bm b_\perp, \bm b_\perp^2)\nonumber\\
&+\frac{\epsilon_\perp^{ij}}{2M}{\partial\over \partial{\bm{b}_{\perp}^j}}\mathcal{H}_{1,2}(x, 0,\bm k_\perp^2, \bm k_\perp\cdot\bm b_\perp, \bm b_\perp^2),
\end{align}
where the distributions $\mathcal{X}$ are the Fourier transformation of the GTMDs $X$:
\begin{align}
\mathcal{X}& = \int {d^2 \bm{\Delta}_\perp \over (2\pi)^2}  e^{-i\bm \Delta_\perp \cdot \bm b_\perp}
X(x, 0,\bm k_\perp^2, \bm k_\perp\cdot\bm \Delta_\perp, \bm \Delta_\perp^2).
\end{align}

\section{Quark Wigner distribution of the pion meson in the overlap representation}

In this section we present the analytical calculation on the quark Wigner distributions of the pion meson in the light-cone quark model using the overlap representation for the parton correlator.
The light-cone formalism has been widely applied in the calculation of parton distribution functions of nucleon and mesons~\cite{Lepage:1980aa,Bacchetta:2008af}.
On the other hand, the overlap representation has also been used to study various form factors of the nucleon~\cite{Brodsky:2000ii} and the pion~\cite{Xiao:2003wf}, anomalous magnetic moment of the nucleon~\cite{Brodsky:2000ii,Lu:2006kt} as well as GPDs~\cite{Brodsky:2000xy}.
Moreover, the light-cone constituent quark model was applied to calculate the quark Wigner distribution of the proton.
It is therefore straightforward to extend the same framework to compute the Wigner distribution of the pion.
In the light-cone approach, the wavefunctions of the hadron, which describe a hadronic composite state at a particular light-cone time, are expressed in terms of a series of light-cone wavefunctions in Fock-state basis.
For example, the Fock states of the pion can be cast into
\begin{align}
|\pi\rangle = |q\bar{q}\rangle + |q\bar{q} g\rangle+\cdots~.
\end{align}
The corresponding light-cone wavefunctions have the form
\begin{eqnarray}
\left\vert \Psi_\pi(P^+, {\bm P_\perp} )\right\rangle &=& \sum_{n}\
\prod_{i=1}^{n}
{{\rm d}x_i\, {\rm d}^2 {\vec k_{\perp i}}
\over \sqrt{x_i}\, 16\pi^3}\ \,
16\pi^3 \delta\left(1-\sum_{i=1}^{n} x_i\right)\,
\nonumber\\
&\times& \delta^{(2)}\left(\sum_{i=1}^{n} {\bm k_{\perp i}}\right)
\psi_n(x_i,{\bm k_{\perp i}},
\lambda_i)
\\
&\times&\left\vert n;\,
x_i P^+, x_i {\bm P_\perp} + {\bm k_{\perp i}}, \lambda_i\right\rangle,
\nonumber
\end{eqnarray}
Where $\lambda_i$ denotes the helicity of parton $i$.
In the following, we only consider the first order contribution in the calculation to simplify the problem, ie., we consider the minimal Fock states of the pion meson ($n=2$), namely, $\psi_\pi(x,\bm{k}_\perp,\lambda_q,\lambda_{\bar{q}})$. These wavefunctions has been derived in Ref.~\cite{Xiao:2003wf} by considering the relativistic effect of quarks~\cite{Melosh:1974cu,Ma:1991xq}:
\begin{align}
\label{eq:LCWFs}
&\psi_{\pi}(x,\bm{k}_\perp,+,-)=+\frac{m}{\sqrt{2(m^2+\bm{k}^2_\perp)}}\varphi_\pi \qquad (l^z=0),\nonumber\\
&\psi_{\pi}(x,\bm{k}_\perp,-,+)=-\frac{m}{\sqrt{2(m^2+\bm{k}^2_\perp)}}\varphi_\pi \qquad (l^z=0),\nonumber\\
&\psi_{\pi}(x,\bm{k}_\perp,+,+)=-\frac{k_{\perp 1}-ik_{\perp 2}}{\sqrt{2(m^2+\bm{k}^2_\perp)}}\varphi_\pi \qquad (l^z=-1),\nonumber\\
&\psi_{\pi}(x,\bm{k}_\perp,-,-)=-\frac{k_{\perp 1}+ik_{\perp 2}}{\sqrt{2(m^2+\bm{k}^2_\perp)}}\varphi_\pi \qquad (l^z=+1).
\end{align}
Here, $x$ is the longitudinal momentum fraction of the quark inside pion, $\bm{k}_\perp = (\bm{k}_{\perp 1}, \bm{k}_{\perp 2} )$ is the intrinsic transverse momentum of the quark, $m$ stands for the mass of the quark/antiquark. We use $+/-$ denotes the positive/negative helicity of the quark and the spectator antiquark. The longitudinal momentum fraction and the intrinsic transverse momentum of the spectator antiquark can be easily determined as $1-x$ and $-\bm  k_\perp$, respectively.
In addition, $\varphi_\pi$ in Eq.~(\ref{eq:LCWFs}) is the wavefunction in momentum space, for which we adopt the Brodsky-Huang-Lepage (BHL) prescription~\cite{Brodsky:1980vj}:
\begin{align}
\label{eq:BHL}
\varphi_\pi(x,\bm{k}_\perp)=A\mathrm{exp}\left[-\frac{1}{8\beta^2}\frac{\bm{k}^2_\perp+m^2}{x(1-x)}\right].
\end{align}

In the overlap representation, the leading-twist generalized correlator can be expressed as
\begin{align}
W^{[\gamma^+]} & ={1\over 16\pi^3}\sum_{\lambda_{\bar{q}}} \left(\psi_{+\lambda_{\bar{q}}}^\star\left(x^{\textrm{out}},\bm k_{\perp}^{\textrm{out}}
 \right)\psi_{+\lambda_{\bar{q}}}\left(x^{\textrm{in}},\bm k_{\perp}^{\textrm{in}}
 \right)\right.\nonumber\\
&+\left.\psi_{-\lambda_{\bar{q}}}^\star\left(x^{\textrm{out}},\bm k_{\perp}^{\textrm{out}}
 \right)\psi_{-\lambda_{\bar{q}}}\left(x^{\textrm{in}},\bm k_{\perp}^{\textrm{in}}
 \right) \right).
\label{eq:wgp} \\
W^{[\gamma^+ \gamma_5]} & = {1\over 16\pi^3}\sum_{\lambda_{\bar{q}}}  \left(\psi_{+\lambda_{\bar{q}}}^\star\left(x^{\textrm{out}},\bm k_{\perp}^{\textrm{out}}
 \right)\psi_{+\lambda_{\bar{q}}}\left(x^{\textrm{in}},\bm k_{\perp}^{\textrm{in}}
 \right)\right.\nonumber\\
&-\left.\psi_{-\lambda_{\bar{q}}}^\star\left(x^{\textrm{out}},\bm k_{\perp}^{\textrm{out}}
 \right)\psi_{-\lambda_{\bar{q}}}\left(x^{\textrm{in}},\bm k_{\perp}^{\textrm{in}}
 \right) \right).
\label{eq:wg5}\\
W^{[i\sigma^{j+}\gamma_5]} &= {1\over 16\pi^3}\sum_{\lambda_{\bar{q}}}  \left(\psi_{\uparrow\lambda_{\bar{q}}}^\star\left(x^{\textrm{out}},\bm k_{\perp}^{\textrm{out}}
 \right)\psi_{\uparrow\lambda_{\bar{q}}}\left(x^{\textrm{in}},\bm k_{\perp}^{\textrm{in}}
 \right)\right.\nonumber\\
&-\left.\psi_{\downarrow\lambda_{\bar{q}}}^\star\left(x^{\textrm{out}},\bm k_{\perp}^{\textrm{out}}
 \right)\psi_{\downarrow\lambda_{\bar{q}}}\left(x^{\textrm{in}},\bm k_{\perp}^{\textrm{in}}
 \right) \right).
\label{eq:wsg5}
\end{align}
where the subscripts $\uparrow$ and $\downarrow $ denote the transverse polarizations of the quark along the direction $\hat{\bm{e}}_x$ and $- \hat{\bm{e}}_x$, respectively.

The arguments of the initial-state wavefunctions in Eqs.~(\ref{eq:wgp}), (\ref{eq:wg5}) and (\ref{eq:wsg5}) are given by
\begin{align}
x^{\textrm{in}}&={x_q-\xi/2\over 1- \xi/2},\nonumber\\
k_{\perp}^{\textrm{in}}&=k_{q\perp} -{1-x_q\over 1-\xi/2}{\bm \Delta_\perp\over 2},  \nonumber
\end{align}
and those of the final-state wavefunctions have the form
\begin{align}
x^{\textrm{out}}&={x_q+\xi/2\over 1+ \xi/2},\nonumber\\
k_{\perp}^{\textrm{out}}&=k_{q\perp} +{1-x_q\over 1+\xi/2}{\bm \Delta_\perp\over 2},  \nonumber
\end{align}

Using the light-cone wavefunctions of the pion in Eq.~(\ref{eq:LCWFs}) as well as the overlap representation for $W^{[\Gamma]}$, we obtain the analytic results of the GTMDs of the pion as follows (in the region $0<\xi/2<x$):
\begin{widetext}
\begin{align}
F_{1,1} & =
C\left(2\bm k_\perp^2 -{(1-x)^2\over (1-\xi^2/4)}{\bm \Delta_\perp^2\over 2}
-{\xi(1-x)\over 1-\xi^2/4} \bm \Delta_\perp \cdot \bm k_\perp+2m^2\right) \nonumber\\
&\times\exp\left( (2x(1+\xi^2/4)-\xi^2)(\bm k_\perp^2 + m^2) +x(1-x)^2\bm \Delta_\perp^2/2-\xi(1-x)^2 \bm k_\perp \cdot \bm\Delta_\perp \over 8\beta^2(x^2-\xi^2/4)(1-x)\right), \\
G_{1,1} & =
-C{2(1-x)M^2\over 1-\xi^2/4}\exp\left( (2x(1+\xi^2/4)-\xi^2)(\bm k_\perp^2 + m^2) +x(1-x)^2\bm \Delta_\perp^2/2-\xi(1-x)^2 \bm k_\perp \cdot \bm\Delta_\perp \over 8\beta^2(x^2-\xi^2/4)(1-x)\right), \\
H_{1,1} & =0,\\
H_{1,2} & =
-C{2(1-x)mM\over 1-\xi^2/4}\exp\left( (2x(1+\xi^2/4)-\xi^2)(\bm k_\perp^2 + m^2) +x(1-x)^2\bm \Delta_\perp^2/2-\xi(1-x)^2 \bm k_\perp \cdot \bm\Delta_\perp \over 8\beta^2(x^2-\xi^2/4)(1-x)\right),
\end{align}
\end{widetext}
where
\begin{align}
C={A^2\over 16\pi^3 B_+ B_-},
\end{align}
with
\begin{align}
B_+& =\sqrt{2\left(\left(\bm k_\perp + \tfrac{1-x}{1+\xi/2}\tfrac{\bm \Delta_\perp^2}{2}\right)^2+ m^2\right)},\\
B_- &=\sqrt{2\left(\left(\bm k_\perp - \tfrac{1-x}{1-\xi/2}\tfrac{\bm \Delta_\perp^2}{2}\right)^2+ m^2\right)}.
\end{align}
In the above result we find that $H_{1,1}$ vanishes in our model, and $H_{1,2}$ is proportional to $G_{1,1}$.

Similarly, the Wigner distribution of the unpolarized quark inside the pion meson can be calculated from the light-cone quark model as
\begin{align}
\rho_{UU}& = \int {d^2 \bm{\Delta}_\perp \over (2\pi)^2}  e^{-i\bm \Delta_\perp \cdot \bm b_\perp}
F_{1,1}(x, 0,\bm k_\perp^2, \bm k_\perp\cdot\bm \Delta_\perp, \bm \Delta_\perp^2) \nonumber\\
&= {1\over 16\pi^3}\sum_{\lambda_{\bar{q}}} \int {d^2 \bm{\Delta}_\perp \over (2\pi)^2}  e^{-i\bm \Delta_\perp \cdot \bm b_\perp}\nonumber\\
 &\times \left(\psi_{+\lambda_{\bar{q}}}^\star\left(x,\bm k_\perp+\tfrac{\bm \Delta_\perp}{ 2}\right)\psi_{+\lambda_{\bar{q}}}\left(x,\bm k_\perp-\tfrac{\bm \Delta_\perp}{ 2}\right)\right.\nonumber\\
 &\left.+\psi_{-\lambda_{\bar{q}}}^\star\left.(x,\bm k_\perp+\tfrac{\bm \Delta_\perp}{ 2}\right) \psi_{-\lambda_{\bar{q}}}\left(x,\bm k_\perp-\tfrac{\bm \Delta_\perp}{ 2}\right)  \right).
\end{align}
where the overlap representation is also applied.
In the same way, one can write down the Wigner distribution of the longitudinally polarized quark inside the pion meson
\begin{align}
\rho_{UL}& = {\partial\over \partial{\bm{b}_{\perp}^i}}\int {d^2 \bm{\Delta}_\perp \over (2\pi)^2}  e^{-i\bm \Delta_\perp \cdot \bm b_\perp}
\frac{i\varepsilon_\perp^{ij} \Delta_\perp^j}{M^2} G_{1,1} \nonumber\\
& =
 {1\over 16\pi^3}\sum_{\lambda_{\bar{q}}} \int {d^2 \bm{\Delta}_\perp \over (2\pi)^2}  e^{-i\bm \Delta_\perp \cdot \bm b_\perp}\nonumber\\
 &\times\left(\psi_{+\lambda_{\bar{q}}}^\star\left(x,\bm k_\perp+\tfrac{\bm \Delta_\perp}{ 2}\right)\psi_{+\lambda_{\bar{q}}}\left(x,\bm k_\perp-\tfrac{\bm \Delta_\perp}{ 2}\right)\right.\nonumber\\
&\left.-\psi_{-\lambda_{\bar{q}}}^\star\left.(x,\bm k_\perp+\tfrac{\bm \Delta_\perp}{ 2}\right) \psi_{-\lambda_{\bar{q}}}\left(x,\bm k_\perp-\tfrac{\bm \Delta_\perp}{ 2}\right)  \right),
\end{align}
as well as the Wigner distribution of the transversely polarized quark inside the pion meson
\begin{align}
\rho_{UT}& =
 {1\over 16\pi^3}\sum_{\lambda_{\bar{q}}} \int {d^2 \bm{\Delta}_\perp \over (2\pi)^2}  e^{-i\bm \Delta_\perp \cdot \bm b_\perp}\nonumber\\
 &\times\left(\psi_{\uparrow\lambda_{\bar{q}}}^\star\left(x,\bm k_\perp+{\bm \Delta_\perp\over 2}\right)\psi_{\uparrow\lambda_{\bar{q}}}\left(x,\bm k_\perp-{\bm{\Delta}_\perp\over 2}\right)\right.\nonumber\\
&\left.-\psi_{\downarrow\lambda_{\bar{q}}}^\star\left.(x,\bm k_\perp+{\bm \Delta_\perp\over 2}\right) \psi_{\downarrow\lambda_{\bar{q}}}\left(x,\bm k_\perp-{\bm \Delta_\perp\over 2}\right)  \right).
\end{align}

\begin{figure*}
  \centering
  \includegraphics[width=0.9\columnwidth]{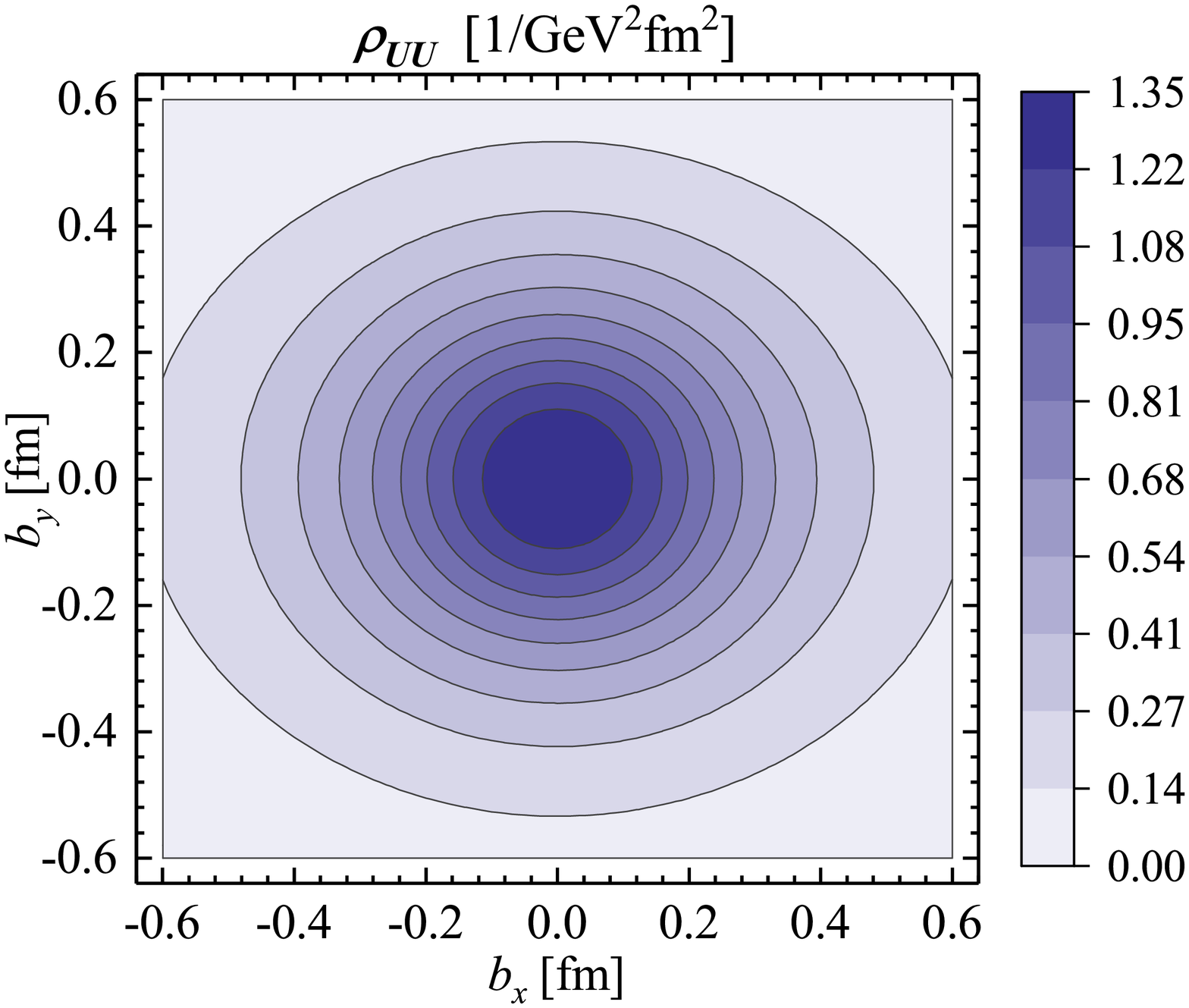}~~~~~~~
  \includegraphics[width=0.9\columnwidth]{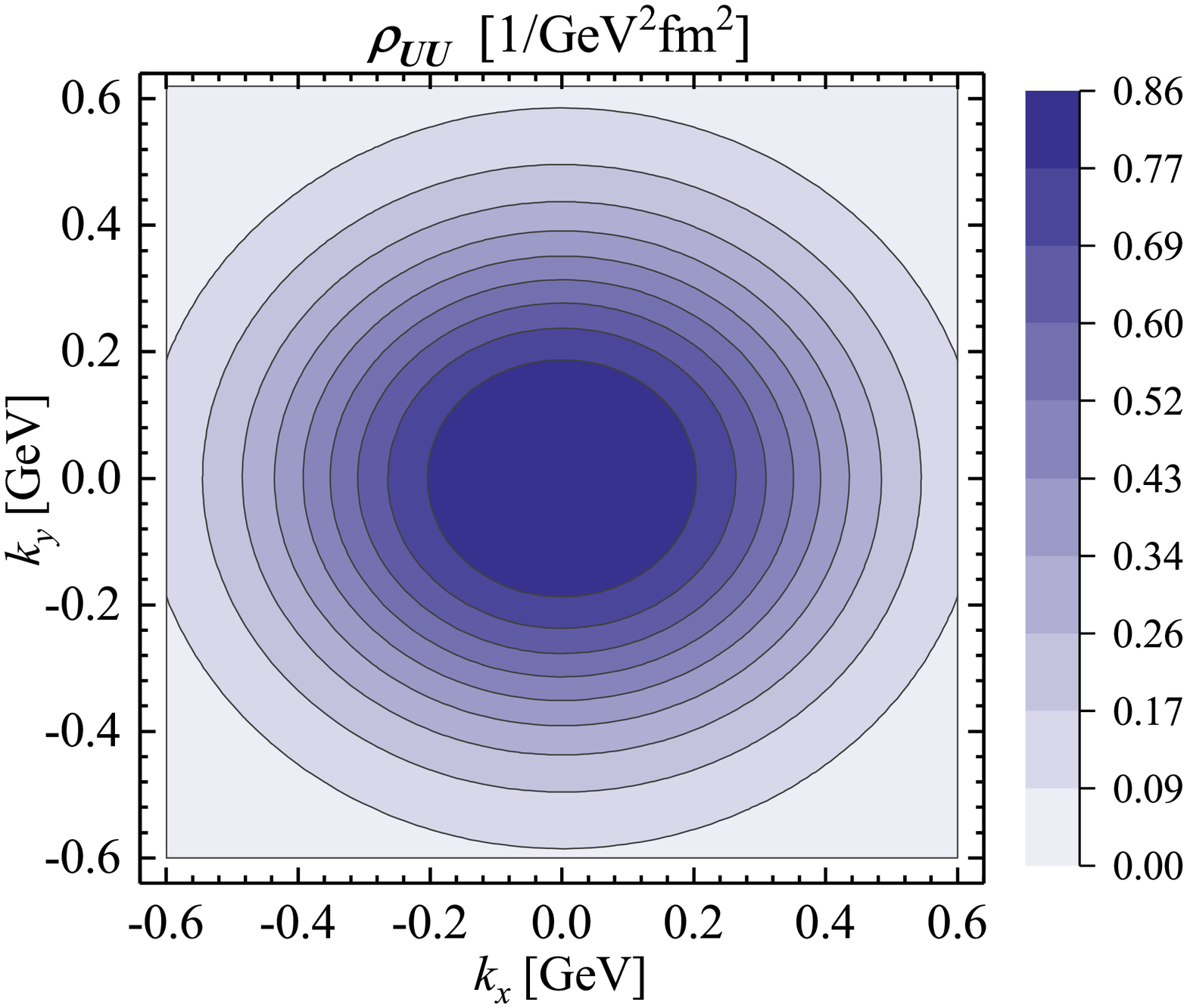}\\
    \includegraphics[width=0.9\columnwidth]{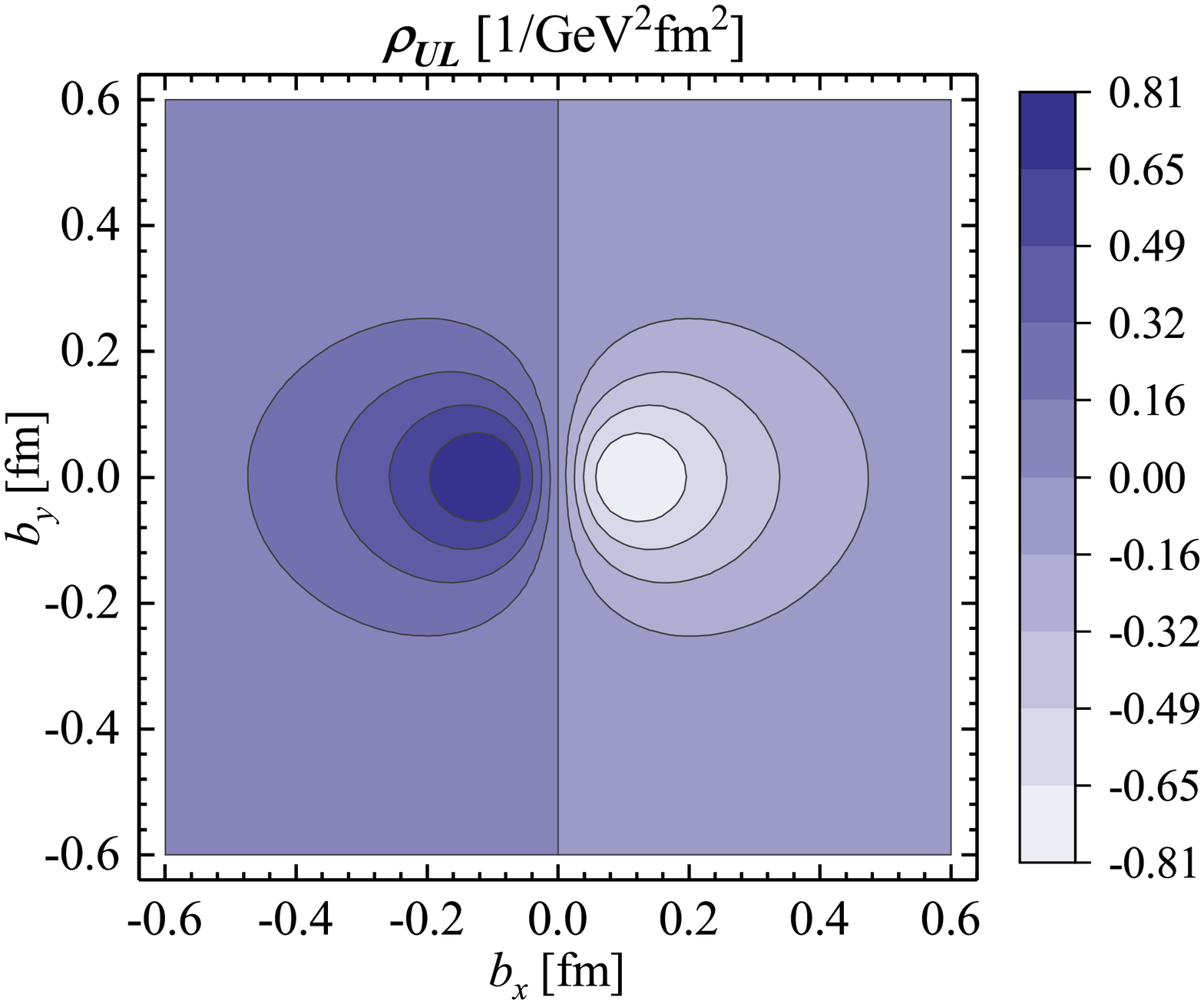}~~~~~~~
  \includegraphics[width=0.9\columnwidth]{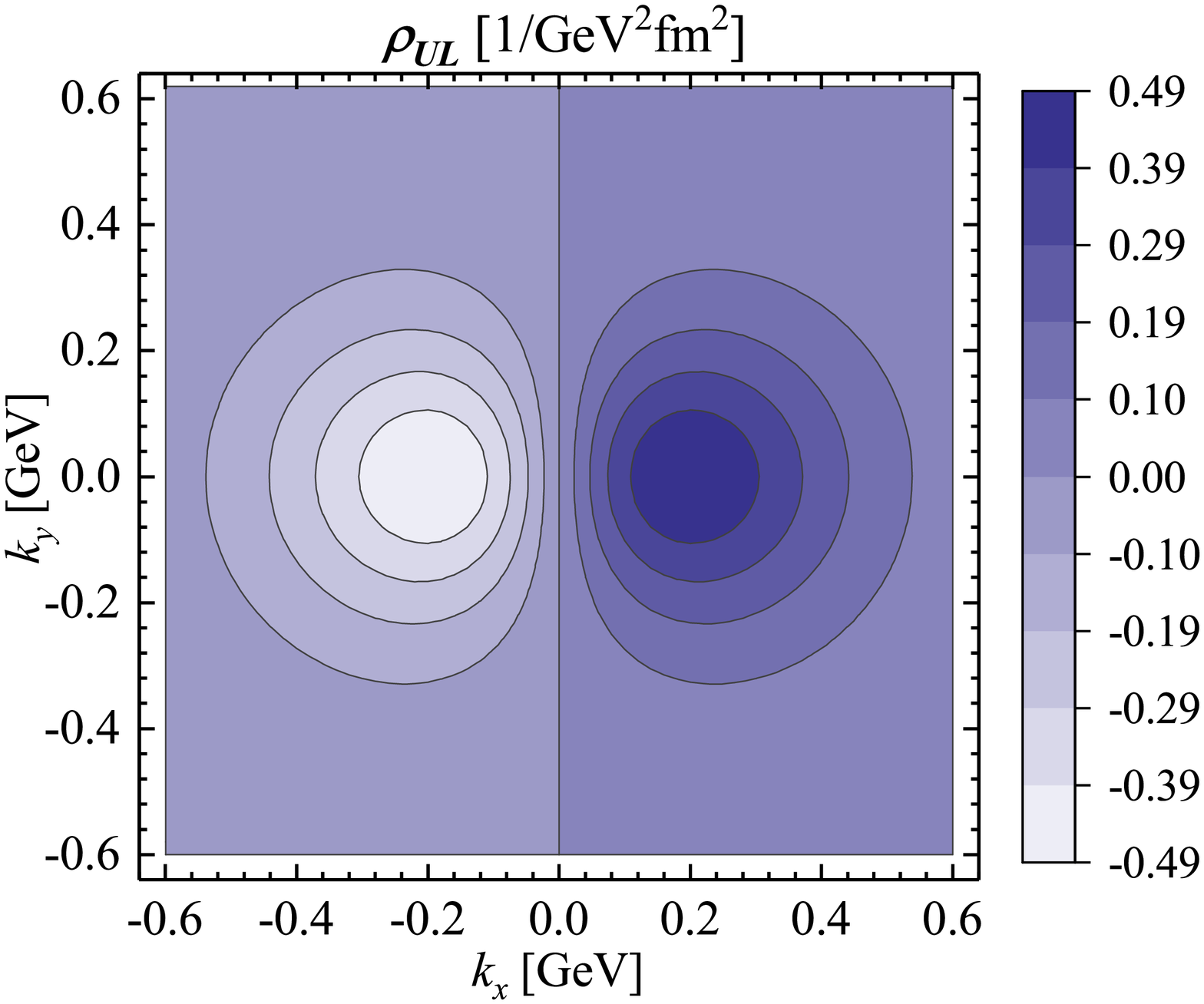}\\
    \includegraphics[width=0.9\columnwidth]{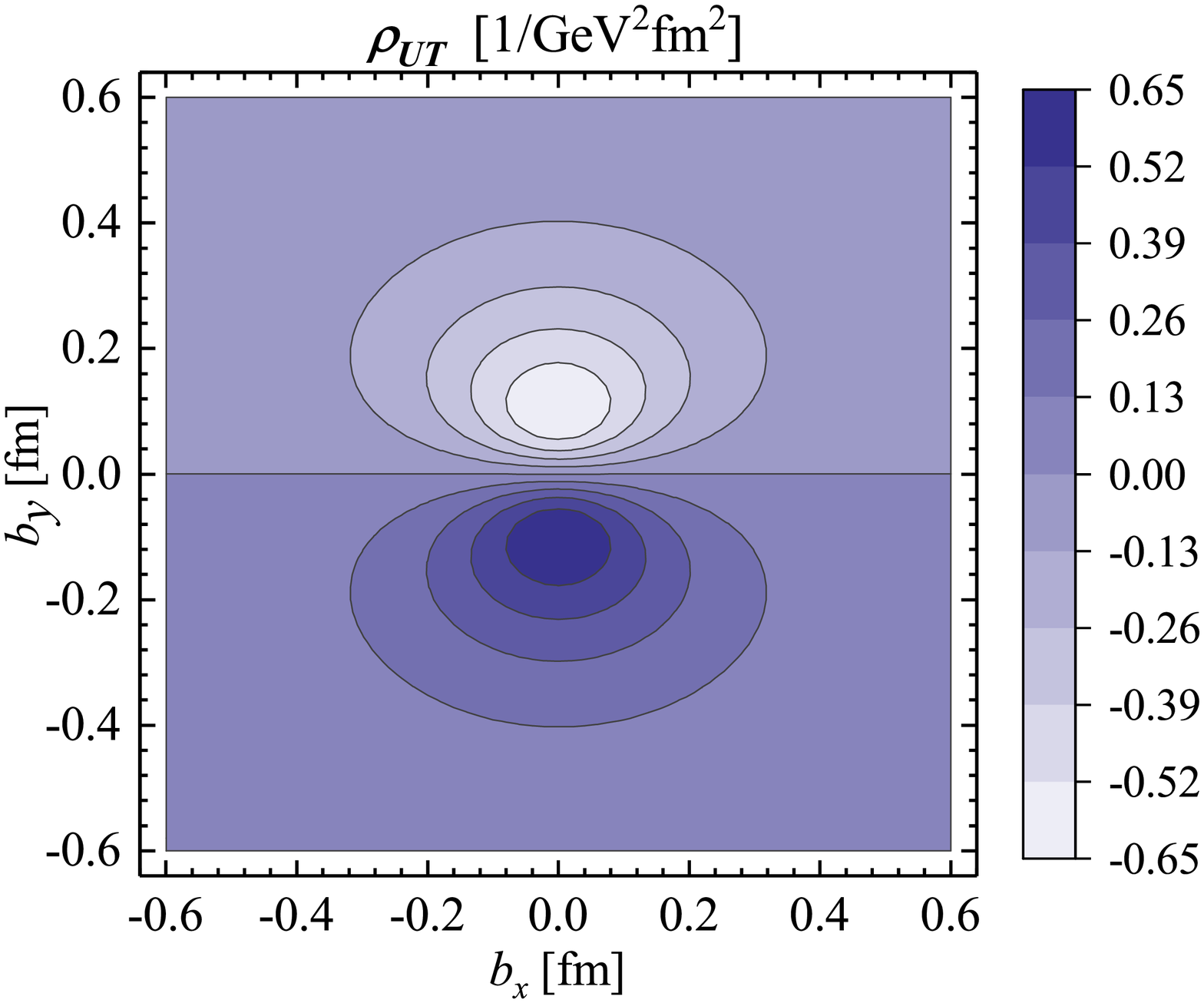}~~~~~~~
  \includegraphics[width=0.9\columnwidth]{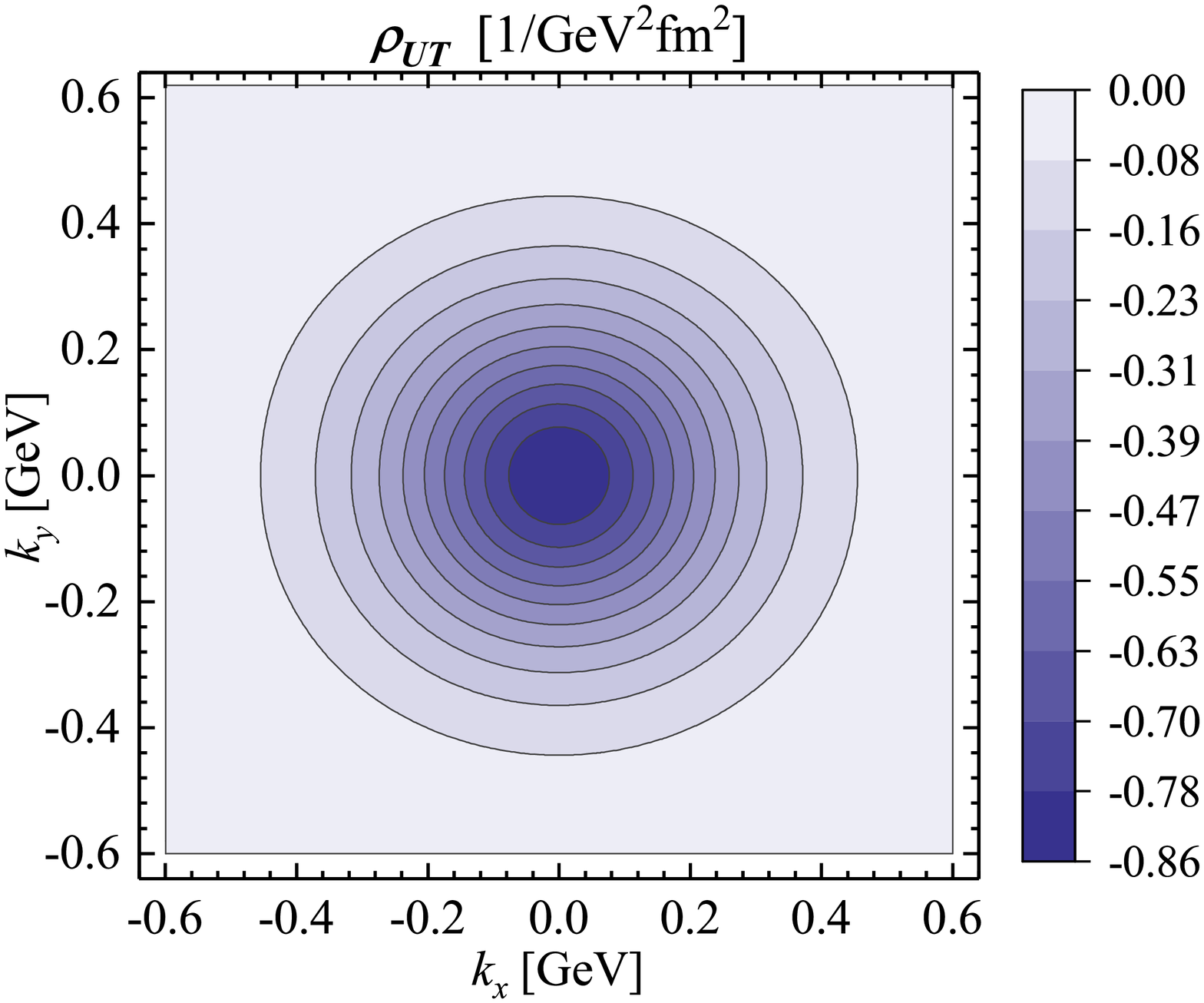}
  \caption{The transverse Wigner distributions $\rho(\bm b_\perp,\bm k_\perp)$ of the pion meson. The upper, central and lower panels show the wigner distributions of the unpolarized quark, the longitudinal polarized quark and the transversely polarized quark inside the pion meson, respectively. the Left panels plot the distribution in the impact-parameter space with fixed transverse momentum $\bm k_\perp = k_\perp \hat{\bm e}_y$ and $k_\perp = 0.3 $ GeV, while the right panels plot the distribution in the transverse-momentum space with fixed impact parameter $\bm b_\perp = b_\perp \hat{\bm e}_y$ and $b_\perp = 0.3$ fm.}
  \label{fig:transbk}
\end{figure*}

Substituting the light-cone wavefunctions of the pion meson in Eq.~(\ref{eq:LCWFs}),
we obtain the expressions for the Wigner distributions in our model as follows:
\begin{widetext}
\begin{align}
\rho_{UU}(x,\bm b_\perp,\bm k_\perp )&=
 {A^2\over 16\pi^3}\int {d^2 \bm{\Delta}_\perp \over (2\pi)^2}  e^{-i\bm \Delta_\perp \cdot \bm b_\perp}\frac{\bm k_\perp^2 -(1-x)^2\bm \Delta_\perp^2/4 +m^2}{\sqrt{\omega_+ \,\omega_-} } \exp\left( -\frac{\bm k_\perp^2 +(1-x)^2\bm \Delta_\perp^2/4+m^2}{ 4\beta^2x(1-x)}\right),\\
\rho_{UL}(x,\bm b_\perp,\bm k_\perp ) &=
 {A^2(1-x)\over 16\pi^3}\int {d^2 \bm{\Delta}_\perp \over (2\pi)^2} e^{-i\bm \Delta_\perp \cdot \bm b_\perp}
\frac{i\epsilon_\perp^{ij}\bm k_{\perp\,i}\bm\Delta_{\perp\,j}}{\sqrt{\omega_+ \,\omega_-} } \exp\left( -\frac{\bm k_\perp^2 +(1-x)^2\bm \Delta_\perp^2/4 +m^2}{ 4\beta^2x(1-x)}\right),\\
\rho_{UT}^j(x,\bm b_\perp,\bm k_\perp ) &=
 {A^2(1-x)\over 16\pi^3}\int {d^2 \bm{\Delta}_\perp \over (2\pi)^2} e^{-i\bm \Delta_\perp \cdot \bm b_\perp}
\frac{im\epsilon_\perp^{ij}\bm\Delta_{\perp i}}{ \sqrt{\omega_+ \,\omega_-} } \exp\left( -\frac{\bm k_\perp^2 + (1-x)^2\bm\Delta_\perp^2/4 +m^2}{ 4\beta^2x(1-x)}\right), \label{eq:rhout}
\end{align}
\end{widetext}
where
\begin{align}
\omega_+ = (\bm k_\perp +(1-x)\tfrac{\bm \Delta_\perp}{ 2})^2+m^2, \nonumber\\
\omega_- = (\bm k_\perp -(1-x)\tfrac{\bm \Delta_\perp}{ 2})^2+m^2.
\end{align}
In this calculation, only the GTMD $H_{1,2}$ contributes to the Wigner distribution $\rho^{[i\sigma^{j+}\gamma_5]}$, since $H_{1,1}$ vanishes in our model.

\section{numerical result for pion Wigner distribution}

To present the numerical result of the pion Wigner distribution, we need to specify the values of the parameters $A$, $m$ and $\beta$ in our model.
We choose the values from Ref.~\cite{Xiao:2003wf,Wang:2017onm}:
\begin{align}
& A=31.303\ \textrm{GeV}^{-1}. \beta=0.41\ \textrm{GeV},\nonumber\\
&m_u=m_d=m=0.2\ \textrm{GeV}.\nonumber
\label{eq:modelscale}
\end{align}

In the following we consider the Wigner distributions in the transverse space, which can be obtained by integrating out the longitudinal momentum fraction $x$:
\begin{align}
\rho^{\Gamma}(\bm b_\perp, \bm k_\perp) = \int_0^1 dx \rho^{\Gamma}(x,\bm b_\perp, \bm k_\perp).
\end{align}

In Fig.~\ref{fig:transbk}, we plot the transverse Wigner distributions $\rho_{UU}(\bm b_\perp, \bm k_\perp)$,
$\rho_{UL}(\bm b_\perp, \bm k_\perp)$ and $\rho_{UT}(\bm b_\perp, \bm k_\perp)$ for the pion meson in the upper, central and lower panels. The Left panels plot the distributions in the impact-parameter space with fixed transverse momentum $\bm k_\perp = k_\perp \hat{\bm e}_y$ and $k_\perp = 0.3$ GeV, while the right panels plot the distributions in the transverse-momentum space with fixed impact parameter $\bm b_\perp = b_\perp \hat{\bm e}_y$ and $b_\perp = 0.3$ fm.

In all the three cases, we find distortions in both the transverse momentum space and the impact parameter space.
However, we observe that the distortion patterns for unpolarized quarks, longitudinally polarized quark and transversely polarized quarks are quite different.
The deformation in $\bm b_\perp$ and $\bm k_\perp$ space of the unpolarized quark in pion meson is similar to result of the valence quarks distributions of the proton~\cite{Lorce:2011kd}, namely, the configuration $b_\perp \perp k_\perp$ is favored.
For the longitudinally polarized quark, positive distribution is observed in the $b_x<0$ region and $k_y>0$ region, this is opposite to the case of the $u$ and $d$ quark distributions for the proton observed in Fig.~4 of Ref.~\cite{Lorce:2011kd} in which the light-cone constituent quark model and the chiral quark-soliton model were adopted. In the $\bm b_\perp$-space, the peak of the distribution of longitudinally polarized quark is around the region $0.1 \,\textrm{fm} <b_x<0.2 \,\textrm{fm}$, while in the $\bm k_\perp$-space the peak is around the region $0.2 \,\textrm{GeV} <k_x<0.3 \,\textrm{GeV}$.
As for the transversely polarized quark with polarization along $\bm e_x$, the distribution is positive in the $b_y<0$ region, while it is negative in the entire $k_\perp$ region.

\begin{figure*}
  \centering
  \includegraphics[width=0.9\columnwidth]{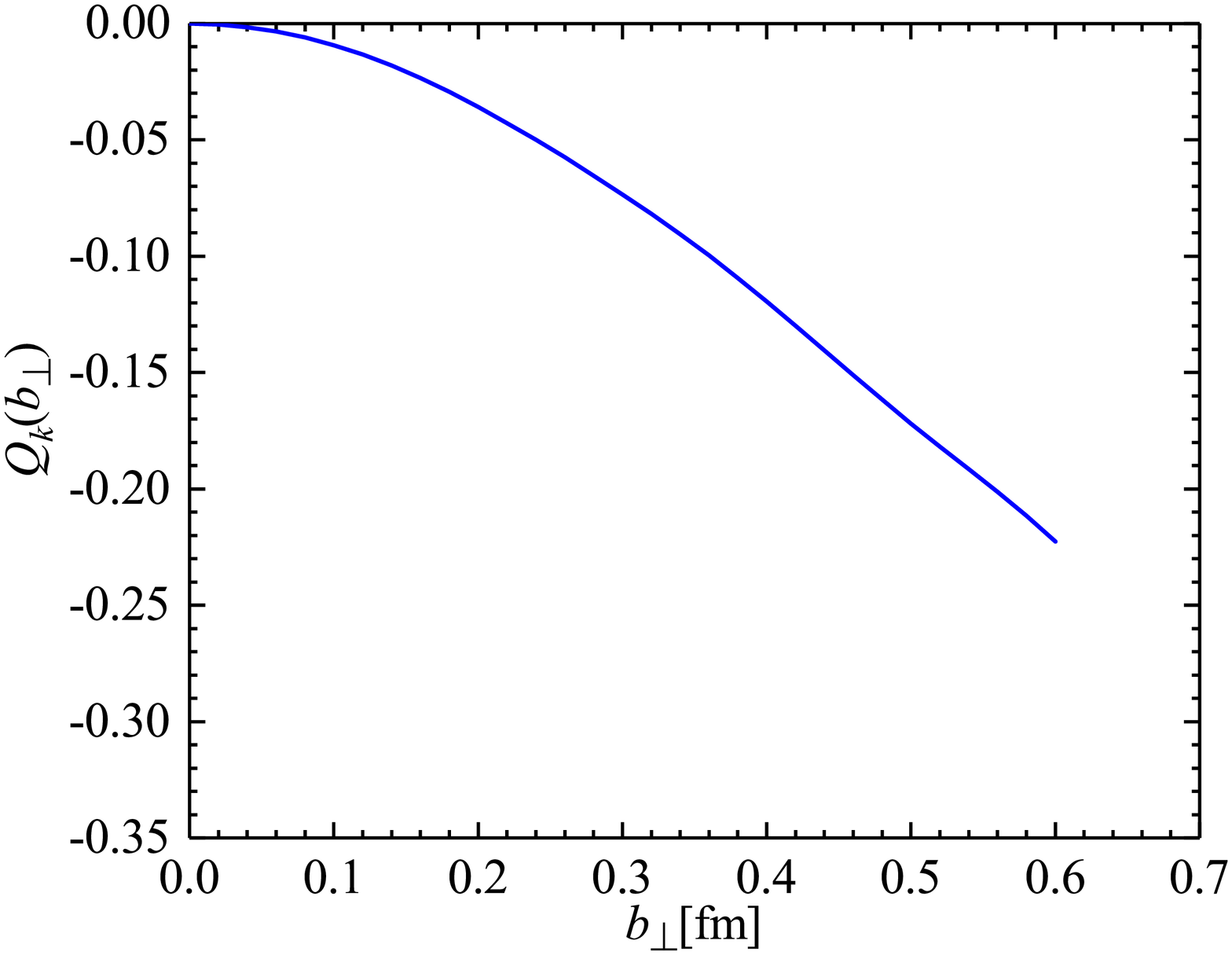}~~~~
  \includegraphics[width=0.9\columnwidth]{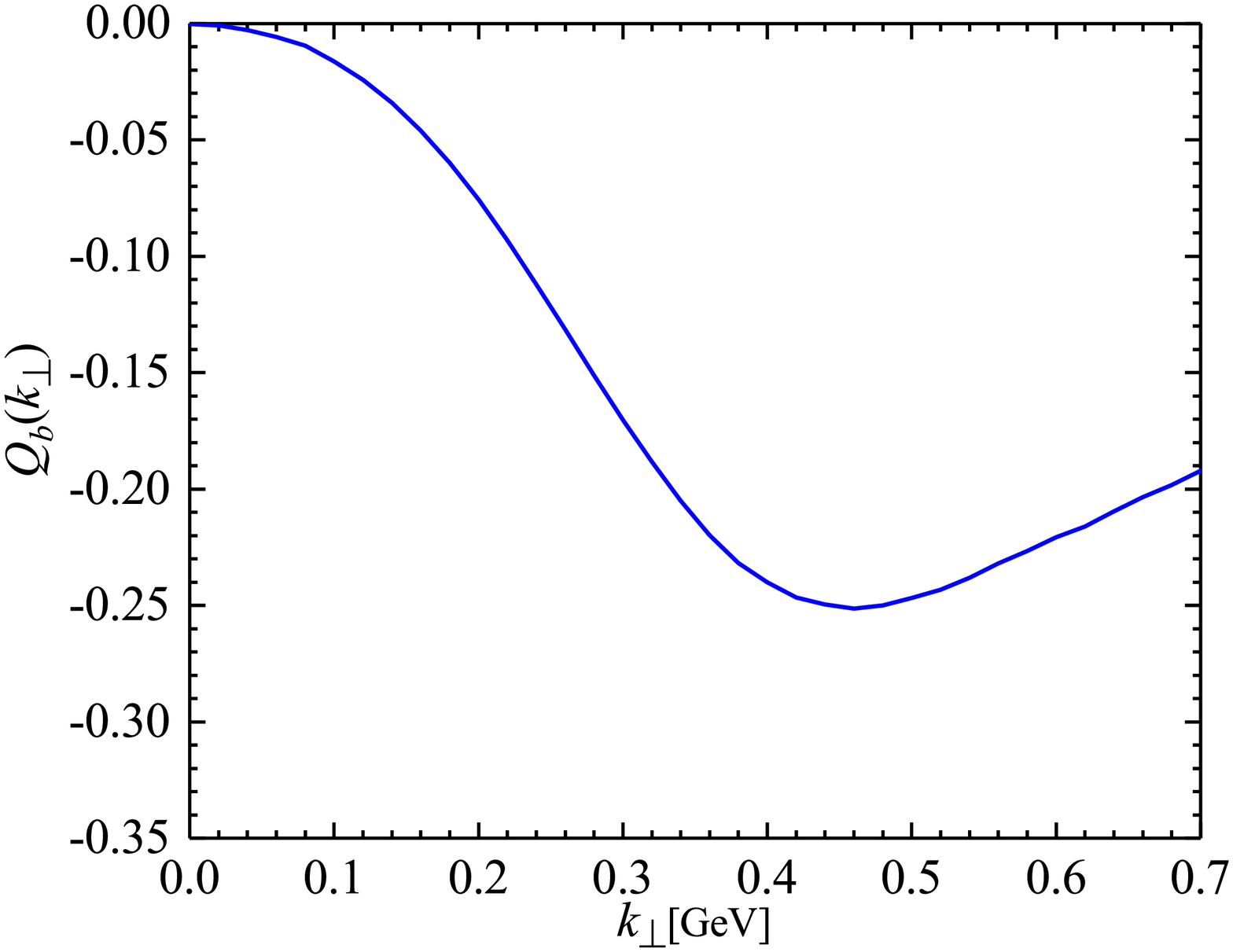}
  \caption{Quadrupole distortions as functions of $b_\perp$ (left panel) and $k_\perp$ (right panel), respectively.}
  \label{fig:quad}
\end{figure*}

To quantitatively estimate the distortion of the unpolarized quark in the pion meson, we also calculate the average quadrupole distortions $Q_b^{ij}(\vec k_\perp)$ and $Q_k^{ij}(\vec b_\perp)$
\begin{align}
Q_b^{ij}(\bm k_\perp)&=Q_b(k_\perp)\left(2\hat{\bm k}_\perp^i\hat{\bm k}^j-\delta^{ij}\right)
\\& =
\frac{\int{\rm d}^2b_\perp\left(2 \bm b^{i}_\perp \bm b^{j}_\perp-\delta^{ij} \bm b^2_\perp\right)\rho_{UU}(\bm b_\perp,\bm k_\perp)}
{\int{\rm d}^2 \bm b_\perp\,b^2_\perp\,\rho_{UU}(\bm b_\perp,\bm k_\perp)},\\
Q_k^{ij}(\bm b_\perp)&=Q_k(b_\perp)\left(2\hat{\bm b}^i\hat{\bm b}^j-\delta^{ij}\right) \\
&=\frac{\int{\rm d}^2\bm k_\perp\left(2\bm k^i_\perp \bm k^j_\perp-\delta^{ij}\bm k^2_\perp\right)
\rho_{UU}(\bm b_\perp,\bm k_\perp)}
{\int{\rm d}^2\bm k_\perp\,k^2_\perp\,\rho_{UU}(\bm b_\perp,\bm k_\perp)},
\label{eq:quadru}
\end{align}
where $\hat{\bm{b}}_\perp = {\bm b_\perp / b_\perp} $ and $\hat{\bm{k}}_\perp = {\bm k_\perp / k_\perp}$.
In the left panel and right panel of Fig.~\ref{fig:quad}, we plot $Q_k(b_\perp)$ vs $b_\perp$ and $Q_b(k_\perp)$ vs $k_\perp$, respectively.
The curves show that the quadrupole distortion for the pion meson is negative, which is the same as that for the proton~\cite{Lorce:2011kd}.
The size of the pion quadrupole distortion is generally smaller than that of proton quadrupole distortion.
We also observe that the $b_\perp$ dependent quadrupole distortion increases with increasing $b_\perp$, while the $b_\perp$ dependent one has the maximum value at $k_\perp=0.45$ GeV.

One can extract more information from the Wigner distribution which is only a quasiprobabilistic density.
This can be done by defining the mixed transverse Wigner distributions~\cite{Lorce:2011kd}
\begin{align}
\tilde{\rho}^{\Gamma}(b_y,k_x) &=\int d b_x \int d k_y \rho^{\Gamma}(\bm b_\perp, \bm k_\perp)
\end{align}
in which $b_x$ and $k_y$ are integrated out.
Different from $\rho^{\Gamma}$, $\tilde{\rho}^{\Gamma}$ has genuine probability interpretation because the remained variables $b_y$ and $k_x$ are commutable.
Therefore, $\tilde{\rho}^{\Gamma}(b_y,k_x)$ represent the probability density of the quark in the $(k_x,b_y)$ plane.
We choose the $(k_x,b_y)$ plane rather than the $(k_y,b_x)$ plane as we find that in our case $\tilde \rho_{UT}(b_x,k_y)$ is consistent with zero.
This is because integrating over $b_y$ leads to the Dirac delta function $\delta(\Delta_y)$ and we adopt the quantized axis of the transverse polarization along $\bm e_x$. Therefore, according to the factor $\epsilon_\perp^{ij} \Delta_i$ in Eq.~\ref{eq:rhout} ($j=x, i=y$),  $\tilde \rho_{UT}(b_x,k_y)$ should vanish.

\begin{figure*}
  \centering
  \includegraphics[width=0.66\columnwidth]{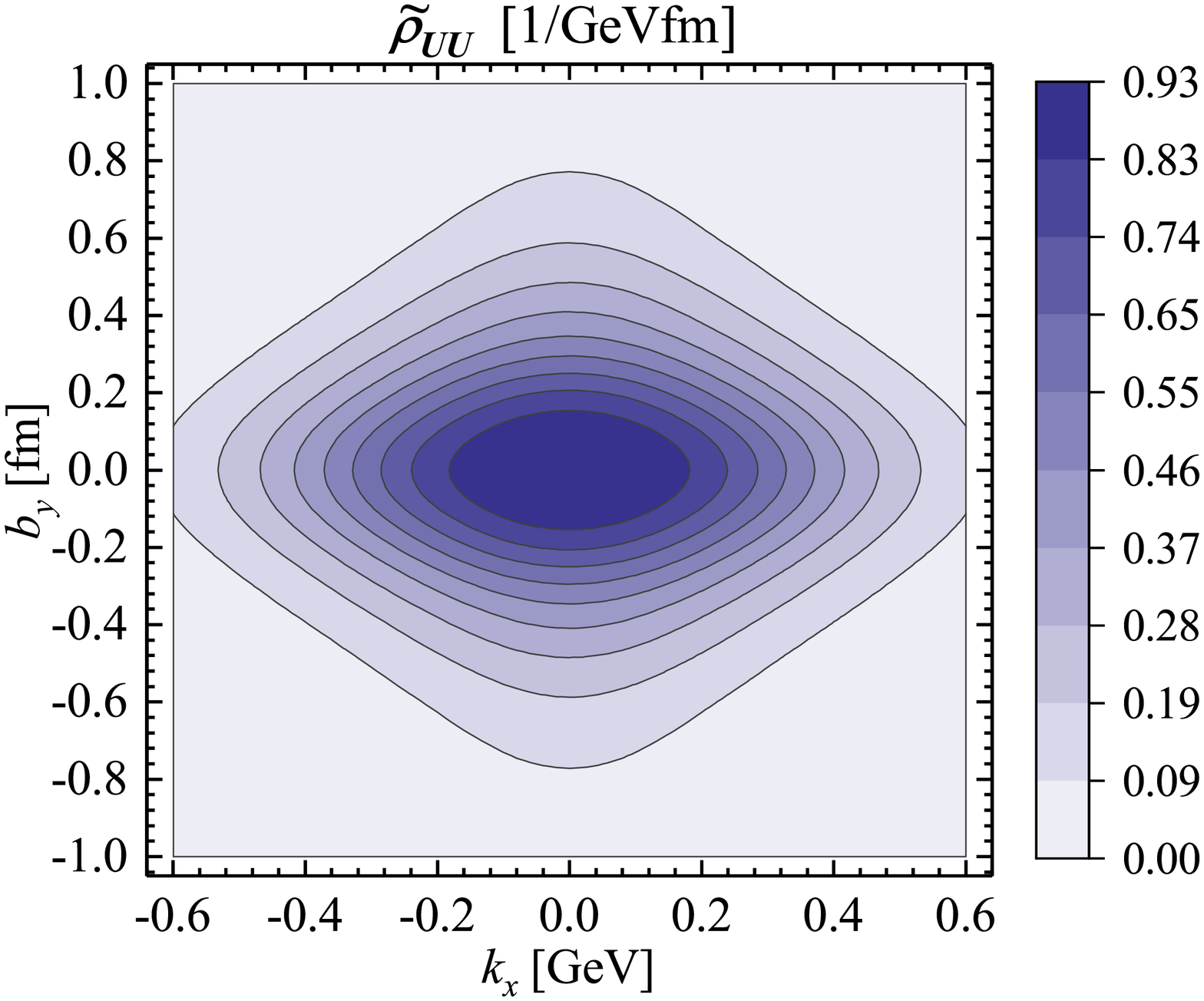}
  \includegraphics[width=0.66\columnwidth]{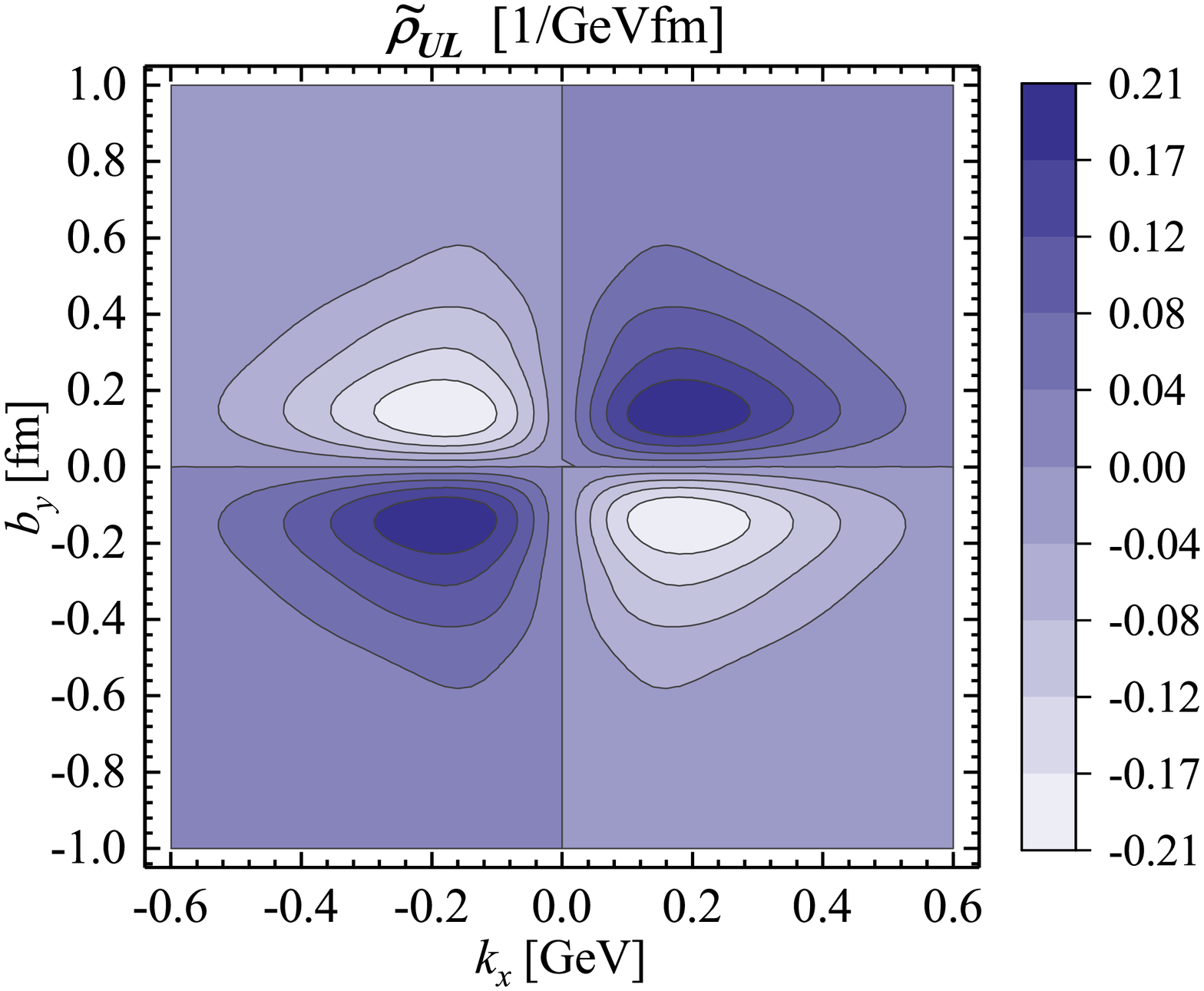}
    \includegraphics[width=0.66\columnwidth]{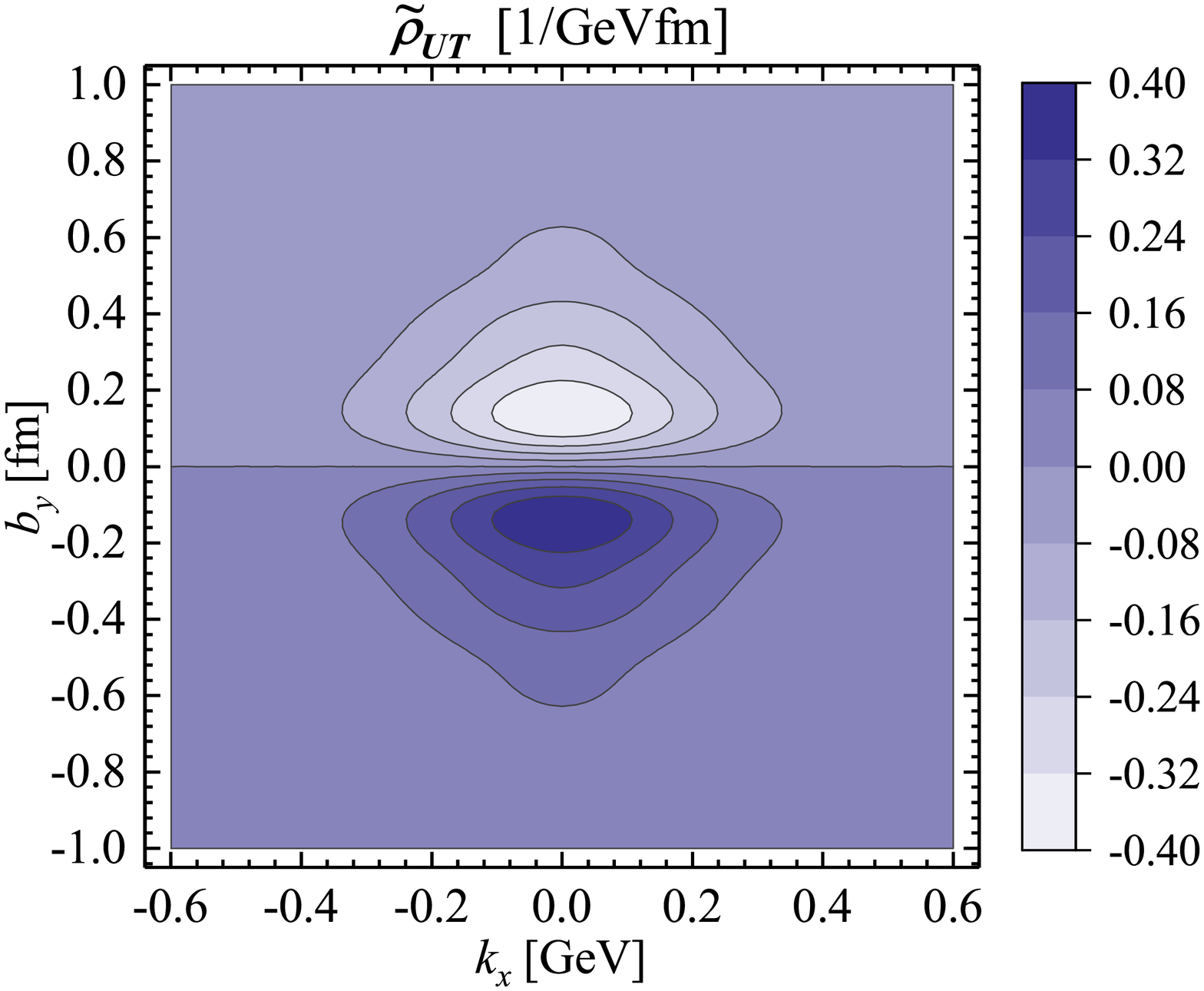}
  \caption{The mixed transverse Wigner distributions $\tilde{\rho}( b_y, k_x)$ for the pion meson. The upper, central and lower panels show the mixed wigner distributions of the unpolarized quark, the longitudinal polarized quark and the transversely polarized quark inside the pion meson, respectively.}
  \label{fig:mixwig}
\end{figure*}

In Fig.~\ref{fig:mixwig}, we plot the mixed transverse density $\tilde \rho_{UU}(b_y,k_x)$, $\tilde \rho_{UL}(b_y,k_x)$ and $\tilde \rho_{UT}(b_y,k_x)$ for the pion meson in the left, central and right panels, respectively.
We find that $\tilde \rho_{UU}(b_y,k_x)$ has the maximum size at $b_y=0\,\textrm{fm}$ and $k_x=0\, \textrm{GeV}$, and it decreases with increasing $b_y$ and $k_x$.
Unlike the unpolarized quarks, the peak of the longitudinally polarized quark distribution $\tilde \rho_{UL}(b_y,k_x)$ is distorted to the region around $|b_y|=0.15\,\textrm{fm}$ and $|k_x|=0.2\, \textrm{GeV}$. The helicity of the quarks tends to be positive in the region $k_x*b_y>0$ and negative in the region $k_x*b_y<0$.
This is similar to the case of the proton Wigner distribution. 
Due to the smaller size of the pion meson, the distortion peaks in the smaller $b_\perp$ region compared with the distortion of the proton.
For the transverse polarization distribution of the quarks along $\bm e_x$, we find that $\tilde \rho_{UT}(b_y,k_x)$ is maximum at $|b_y|=0.15\,\textrm{fm}$ and $|k_x|=0\, \textrm{GeV}$, and the equidensity lines are approximately triangles in the outer region. Also, the transverse polarization of the quark is positive in the $b_y<0$ region and negative in the $b_y>0$ region.

\section{conclusion}

In this work, we studied the quark Wigner distributions of the pion meson using a light-cone quark model.
In the model, the Wigner distributions as well as the GTMDs can be expressed as the overlap of the light-cone wavefunctions of the initial- and final-state pion meson.
We presented the analytic expressions of the transverse Wigner distribution for the unpolarized quark $ \rho_{UU}(\bm b_\perp,\bm k_\perp)$,  longitudinally polarized quark $\rho_{UL}(\bm b_\perp,\bm k_\perp)$ and transversely polarized quark $\rho_{UT}(\bm b_\perp,\bm k_\perp)$, using the model resulting light-cone wavefunctions.
The analytic results for the leading-twist GTMDs $F_{1,1}$, $G_{1,1}$ and $H_{1,2}$ in the region $0<\xi/2<x$ were also given.
By properly chosen the values of the parameters in the model, we plotted the equidensity lines of the transverse Wigner distribution $\rho(\bm b_\perp,\bm k_\perp)$.
we found that there are distortions of quark distributions in all the three cases.
The deformation of the unpolarized quark inside pion meson is similar to that inside proton shown in a previous study, while the deformation of the longitudinally quark inside pion meson is different from that inside the proton.
We also calculated the averaged quadrupole distortions $Q_b^{ij}(\vec k_\perp)$ and $Q_k^{ij}(\vec b_\perp)$ which quantitatively denotes the distortion of unpolarized quark inside pion meson,
Finally, we estimated the mixed Wigner distributions $\tilde \rho_{UU}(b_y,k_x)$, $\tilde \rho_{UL}(b_y,k_x)$ and $\tilde \rho_{UT}(b_y,k_x)$.
The study may provide information of quarks inside the pion meson in the transverse momentum space as well as the impact-parameter space.

\section*{Acknowledgements}
This work is partially supported by the NSFC (China) grant 11575043 and by the Fundamental Research Funds for the Central Universities of China.


\begin{thebibliography}{99}


\bibitem{Feynman:1969ej}
  R.~P.~Feynman,
  Phys.\ Rev.\ Lett.\  {\bf 23}, 1415 (1969).

\bibitem{Mulders:1995dh}
  P.~J.~Mulders and R.~D.~Tangerman,
  Nucl.\ Phys.\ B {\bf 461}, 197 (1996)
  Erratum: [Nucl.\ Phys.\ B {\bf 484}, 538 (1997)]
[hep-ph/9510301].

  \bibitem{Goeke:2005hb}
  K.~Goeke, A.~Metz and M.~Schlegel,
  Phys.\ Lett.\ B {\bf 618}, 90 (2005)
  [hep-ph/0504130].

\bibitem{Bacchetta:2006tn}
  A.~Bacchetta, M.~Diehl, K.~Goeke, A.~Metz, P.~J.~Mulders and M.~Schlegel,
  JHEP {\bf 0702}, 093 (2007)
  [hep-ph/0611265].


\bibitem{Sivers:1989cc}
  D.~W.~Sivers,
  Phys.\ Rev.\ D {\bf 41}, 83 (1990).
\bibitem{Sivers:1990fh}
  D.~W.~Sivers,
  Phys.\ Rev.\ D {\bf 43}, 261 (1991).

\bibitem{Boer:1997nt}
  D.~Boer and P.~J.~Mulders,
  Phys.\ Rev.\ D {\bf 57}, 5780 (1998)
  [hep-ph/9711485].
\bibitem{Brodsky:2002cx}
  S.~J.~Brodsky, D.~S.~Hwang and I.~Schmidt,
  Phys.\ Lett.\ B {\bf 530}, 99 (2002).

\bibitem{Collins:2002kn}
  J.~C.~Collins,
  Phys.\ Lett.\ B {\bf 536}, 43 (2002)
  [hep-ph/0204004].

\bibitem{Ji:2002aa}
  X.~D.~Ji and F.~Yuan,
  Phys.\ Lett.\ B {\bf 543}, 66 (2002)
  [hep-ph/0206057].




\bibitem{Mueller:1998fv}
  D.~Mueller, D.~Robaschik, B.~Geyer, F.~M.~Dittes and J.~Horejsi,
  Fortsch.\ Phys.\  {\bf 42}, 101 (1994)
  [arXiv:hep-ph/9812448].

\bibitem{Ji:1996nm}
  X.~D.~Ji,
  Phys.\ Rev.\  D {\bf 55}, 7114 (1997)
  [arXiv:hep-ph/9609381].

\bibitem{Radyushkin:1997ki}
  A.~V.~Radyushkin,
  Phys.\ Rev.\  D {\bf 56}, 5524 (1997)
  [arXiv:hep-ph/9704207].

\bibitem{Burkardt:2000za}
  M.~Burkardt,
  Phys.\ Rev.\  D {\bf 62}, 071503 (2000)
  [Erratum-ibid.\  D {\bf 66}, 119903 (2002)]
  [arXiv:hep-ph/0005108].

\bibitem{Burkardt:2002hr}
  M.~Burkardt,
  Int.\ J.\ Mod.\ Phys.\  A {\bf 18}, 173 (2003)
  [arXiv:hep-ph/0207047].

\bibitem{Diehl:2002he}
  M.~Diehl,
  Eur.\ Phys.\ J.\  C {\bf 25}, 223 (2002)
  [Erratum-ibid.\  C {\bf 31}, 277 (2003)]
  [arXiv:hep-ph/0205208].

\bibitem{Ji:2003ak}
  X.~d.~Ji,
  Phys.\ Rev.\ Lett.\  {\bf 91}, 062001 (2003)
  [hep-ph/0304037].
\bibitem{Belitsky:2003nz}
  A.~V.~Belitsky, X.~d.~Ji and F.~Yuan,
  Phys.\ Rev.\ D {\bf 69}, 074014 (2004)
  [hep-ph/0307383].

\bibitem{Lorce:2011kd}
  C.~Lorce and B.~Pasquini,
  Phys.\ Rev.\ D {\bf 84}, 014015 (2011)
  [arXiv:1106.0139 [hep-ph]].

\bibitem{Brodsky:1997de}
  S.~J.~Brodsky, H.~C.~Pauli and S.~S.~Pinsky,
  Phys.\ Rept.\  {\bf 301}, 299 (1998)
  [hep-ph/9705477].



  \bibitem{Lorce:2011ni}
  C.~Lorce, B.~Pasquini, X.~Xiong and F.~Yuan,
  Phys.\ Rev.\ D {\bf 85}, 114006 (2012)
  [arXiv:1111.4827 [hep-ph]].


\bibitem{Liu:2015eqa}
  T.~Liu and B.~Q.~Ma,
  Phys.\ Rev.\ D {\bf 91}, 034019 (2015)
  [arXiv:1501.07690 [hep-ph]].

\bibitem{Mukherjee:2014nya}
  A.~Mukherjee, S.~Nair and V.~K.~Ojha,
  Phys.\ Rev.\ D {\bf 90}, no. 1, 014024 (2014)
  [arXiv:1403.6233 [hep-ph]].

\bibitem{Mukherjee:2015aja}
  A.~Mukherjee, S.~Nair and V.~K.~Ojha,
  Phys.\ Rev.\ D {\bf 91}, no. 5, 054018 (2015)
  [arXiv:1501.03728 [hep-ph]].

\bibitem{More:2017zqp}
  J.~More, A.~Mukherjee and S.~Nair,
  Eur.\ Phys.\ J.\ C {\bf 78}, no. 5, 389 (2018)
  [arXiv:1709.00943 [hep-ph]].

\bibitem{Hatta:2016dxp}
  Y.~Hatta, B.~W.~Xiao and F.~Yuan,
  Phys.\ Rev.\ Lett.\  {\bf 116}, no. 20, 202301 (2016)
  [arXiv:1601.01585 [hep-ph]].

\bibitem{Hatta:2017cte}
  Y.~Hatta, B.~W.~Xiao and F.~Yuan,
  Phys.\ Rev.\ D {\bf 95}, no. 11, 114026 (2017)
  [arXiv:1703.02085 [hep-ph]].

\bibitem{Hagiwara:2017fye}
  Y.~Hagiwara, Y.~Hatta, R.~Pasechnik, M.~Tasevsky and O.~Teryaev,
  Phys.\ Rev.\ D {\bf 96}, no. 3, 034009 (2017)
  [arXiv:1706.01765 [hep-ph]].

\bibitem{Bhattacharya:2017bvs}
  S.~Bhattacharya, A.~Metz and J.~Zhou,
  Phys.\ Lett.\ B {\bf 771}, 396 (2017)
  [arXiv:1702.04387 [hep-ph]].

\bibitem{Xiao:2003wf}
  B.~W.~Xiao and B.~Q.~Ma,
  Phys.\ Rev.\ D {\bf 68}, 034020 (2003).

\bibitem{Meissner:2008ay}
  S.~Meissner, A.~Metz, M.~Schlegel and K.~Goeke,
  JHEP {\bf 0808}, 038 (2008)
  [arXiv:0805.3165 [hep-ph]].




\bibitem{Meissner:2009ww}
  S.~Meissner, A.~Metz and M.~Schlegel,
  JHEP {\bf 0908}, 056 (2009)
  [arXiv:0906.5323 [hep-ph]].




\bibitem{Lepage:1980aa} G. P. Lepage and S.
J. Brodsky, Phys. Rev. D 22 (1980) 2157; Phys. Lett. B 87 (1979) 359; Phys.
Rev. Lett. 43 (1979) 545, 1625(E).


\bibitem{Bacchetta:2008af}
  A.~Bacchetta, F.~Conti and M.~Radici,
  Phys.\ Rev.\ D {\bf 78}, 074010 (2008)
  [arXiv:0807.0323 [hep-ph]].











\bibitem{Brodsky:2000ii}
  S.~J.~Brodsky, D.~S.~Hwang, B.~Q.~Ma and I.~Schmidt,
  Nucl.\ Phys.\ {\bf B593}, 311 (2001).


  \bibitem{Lu:2006kt}
  Z.~Lu and I.~Schmidt,
  Phys.\ Rev.\ D {\bf 75}, 073008 (2007)
 [hep-ph/0611158].






\bibitem{Brodsky:2000xy}
  S.~J.~Brodsky, M.~Diehl and D.~S.~Hwang,
  Nucl.\ Phys.\ B {\bf 596}, 99 (2001)
  [hep-ph/0009254].

\bibitem{Melosh:1974cu}
  H.~J.~Melosh,
  Phys.\ Rev.\ D {\bf 9}, 1095 (1974).

\bibitem{Ma:1991xq}
  B.~Q.~Ma,
  J.\ Phys.\ G {\bf 17}, L53 (1991)
  [arXiv:0711.2335 [hep-ph]].


\bibitem{Brodsky:1980vj}%
  S.~J.~Brodsky, T.~Huang and G.~P.~Lepage,
  \href {http://www-public.slac.stanford.edu/sciDoc/docMeta.aspx?slacPubNumber=SLAC-PUB-2540}
        {SLAC-PUB-2540 (1980)};\ %
  S.~J.~Brodsky, T.~Huang\ and\ G.~P.~Lepage,
  \href {} {Conf. Proc.\ \textbf {C810816},\ 143 (1981)};\ %
  S.~J.~Brodsky, T.~Huang and G.~P.~Lepage,\
  \href {http://www-public.slac.stanford.edu/sciDoc/docMeta.aspx?slacPubNumber=SLAC-PUB-2868}
        {Springer Tracts Mod.\ Phys.\ \textbf {100}, 81 (1982)}.%








\bibitem{Wang:2017onm}
  Z.~Wang, X.~Wang and Z.~Lu,
  Phys.\ Rev.\ D {\bf 95}, 094004 (2017)
  [arXiv:1702.03637 [hep-ph]].
\end{thebibliography}
\end{document}